\documentclass[sigconf,screen,nonacm]{acmart}
\renewcommand\footnotetextcopyrightpermission[1]{} %remove copyright
\AtBeginDocument{%
  }

\usepackage{eso-pic}

\newcommand{\myACMFooter}{%
  \AddToShipoutPictureFG*{%
    \AtPageLowerLeft{%
      \hspace*{0.75in}% 左边距，可微调
      \raisebox{0.6in}{% 距离页面底部，可微调
        \begin{minipage}{0.42\textwidth}
          \rule{\linewidth}{0.4pt}\\[-0.2em]
          {\footnotesize
            Submitted to Conference '26, USA.
            Manuscript under review.
          }
        \end{minipage}
      }%
    }%
  }%
}

\usepackage{makecell}
\usepackage{soul}
\usepackage{pifont}
\usepackage{float}      
\usepackage{placeins} 
\usepackage{capt-of} 
\usepackage{multicol}
\usepackage{multirow}
\usepackage{colortbl}
\usepackage{pifont}
\usepackage{graphicx}
\usepackage{subcaption}
\usepackage{etoolbox}
\usepackage{hyperref}
\usepackage{amsmath}
\usepackage{geometry}
\usepackage{booktabs}   
\usepackage{makecell}
\usepackage{enumitem}
\usepackage{amssymb}
\usepackage{textcomp}
\usepackage{array}
\usepackage{xeCJK}
\usepackage{stfloats}

\begin{document}

%%
%% The "title" command has an optional parameter,
%% allowing the author to define a "short title" to be used in page headers.
\title[Exascale Hybrid Numerical-AI Ensembles for Operational Flood-Season Forecasting in East Asia]{Exascale Hybrid Numerical-AI Ensembles for Operational Flood-Season Forecasting in East Asia: 15-km Decadal Hindcasts and 1-km High-Resolution Capability}

\author{
  \begin{tabular}{@{}c@{}}
    {Mengxuan Chen}$^{1,*}$, {Yunpu Xu}$^{1,*}$, {Qiuyan Sun}$^{1,*}$, {Han Zhang}$^{2,3}$, {Jiayi Lai}$^{4}$, {Zheng Zhou}$^{1}$, \\
    {Juepeng Zheng}$^{2,7,}$\textsuperscript{\textdagger}, {Hongsong Meng}$^{6}$, {Nan Wei}$^{2,}$\textsuperscript{\textdagger}, {Jinxiao Zhang}$^{1}$, 
    {Xiongchuan Tan}$^{1}$, \\
    {Haodong Bian}$^{5}$, {Yinan Cai}$^{7}$, {Ge Yang}$^{7}$, {Fang Wang}$^{8,}$\textsuperscript{\textdagger}, {Yunyun Liu}$^{8}$, {Conghui He}$^{9}$, \\
    {Runmin Dong}$^{2}$, {Lanning Wang}$^{4}$, {Yutong Lu}$^{2,7}$, {Yongjiu Dai}$^{2}$, {Haohuan Fu}$^{1,6,7,}$\textsuperscript{\textdagger}
  \end{tabular}
}
\affiliation{
  \begin{tabular}{@{}c@{}}
    $^{1}$Tsinghua University \quad $^{2}$Sun Yat-Sen University \quad $^{3}$Jiangsu Provincial Meteorol. Bureau \quad $^{4}$Beijing Normal University\\ 
    $^{5}$Qinghai University \quad $^{6}$National Supercomputing Center in Wuxi \quad $^{7}$National Supercomputing Center in Shenzhen\\ 
    $^{8}$CMA Earth System Modeling and Prediction Center \quad $^{9}$Shanghai Artificial Intelligence Laboratory\\
    $^{*}$Equal contribution, \textsuperscript{\textdagger}Corresponding author 
  \end{tabular}
  \country{}
}

%%
%% By default, the full list of authors will be used in the page
%% headers. Often, this list is too long, and will overlap
%% other information printed in the page headers. This command allows
%% the author to define a more concise list
%% of authors' names for this purpose.
% \makeatletter
% \def\@shortauthors{}
% \makeatother
\renewcommand{\shortauthors}{Chen, M., Xu, Y., Sun, Q., et al.}

%%
%% The abstract is a short summary of the work to be presented in the
%% article.
\begin{abstract}
Seasonal forecasting of summer rainfall in East Asia remains a grand challenge, as predictability at 3 to 6 month lead times is constrained by the spring predictability barrier, weak large-scale signals, and localized nonlinear convective extremes. We address this challenge with CAPES, which integrates a kilometer-resolution coupled regional model with atmosphere, land, and ocean components and a data-driven AI seasonal forecasting system. At 15 km resolution, the fused workflow combines 174 numerical members from varying start times, physics schemes, and parameter perturbations with 1,600 AI members generated from initial-state and model-latent perturbations. Using the full LineShine system, CAPES completes ten annual 1,774-member hindcasts for 2016 to 2025 within 14.6 hours, improving the mean prediction score from ECMWF's 71.8 to 75.9 and delivering a major gain in operational forecasting capability. The 1-km configuration further enables fine-scale typhoon simulation and establishes the feasibility of kilometer-scale fused ensemble forecasting on a one-week timescale.
\end{abstract}

%%
%% The code below is generated by the tool at http://dl.acm.org/ccs.cfm.
%% Please copy and paste the code instead of the example below.
%%
% \begin{CCSXML}
% <ccs2012>
%  <concept>
%   <concept_id>00000000.0000000.0000000</concept_id>
%   <concept_desc>Do Not Use This Code, Generate the Correct Terms for Your Paper</concept_desc>
%   <concept_significance>500</concept_significance>
%  </concept>
%  <concept>
%   <concept_id>00000000.00000000.00000000</concept_id>
%   <concept_desc>Do Not Use This Code, Generate the Correct Terms for Your Paper</concept_desc>
%   <concept_significance>300</concept_significance>
%  </concept>
%  <concept>
%   <concept_id>00000000.00000000.00000000</concept_id>
%   <concept_desc>Do Not Use This Code, Generate the Correct Terms for Your Paper</concept_desc>
%   <concept_significance>100</concept_significance>
%  </concept>
%  <concept>
%   <concept_id>00000000.00000000.00000000</concept_id>
%   <concept_desc>Do Not Use This Code, Generate the Correct Terms for Your Paper</concept_desc>
%   <concept_significance>100</concept_significance>
%  </concept>
% </ccs2012>
% \end{CCSXML}

% \ccsdesc[500]{Do Not Use This Code~Generate the Correct Terms for Your Paper}
% \ccsdesc[300]{Do Not Use This Code~Generate the Correct Terms for Your Paper}
% \ccsdesc{Do Not Use This Code~Generate the Correct Terms for Your Paper}
% \ccsdesc[100]{Do Not Use This Code~Generate the Correct Terms for Your Paper}

%%
%% Keywords. The author(s) should pick words that accurately describe
%% the work being presented. Separate the keywords with commas.

%% A "teaser" image appears between the author and affiliation
%% information and the body of the document, and typically spans the
%% page.
% \begin{teaserfigure}
%   \includegraphics[width=\textwidth]{sampleteaser}
%   \caption{Seattle Mariners at Spring Training, 2010.}
%   \Description{Enjoying the baseball game from the third-base
%   seats. Ichiro Suzuki preparing to bat.}
%   \label{fig:teaser}
% \end{teaserfigure}

% \received{20 February 2007}
% \received[revised]{12 March 2009}
% \received[accepted]{5 June 2009}

%%
%% This command processes the author and affiliation and title
%% information and builds the first part of the formatted document.
% \makeatletter
% \def\@shortauthors{}
\maketitle
\enlargethispage{1cm}  % 增大第一页下边距
\myACMFooter
%\pagestyle{plain}

% https://awards.acm.org/bell/nominations#h-submissions
\section{Justification for ACM Gordon Bell Prize For Climate Modeling}
% (50 word max)
% indicate what implementation or performance “high watermark” was achieved (rather than the science that was enabled)
We convert full-system exascale computing into operational flood-season forecasting capability by integrating large-scale numerical simulation and AI ensemble generation, executing ten annual 1,774-member hybrid hindcasts within 14.6 hours, and improving the mean prediction score from ECMWF's 71.8 to 75.9.

\section{Performance Attributes}
\noindent\makebox[\linewidth][c]{%
\resizebox{1\linewidth}{!}{%
    \begin{tabular}{|c|c|}
    \hline
    Attributes & Contents \\
    \hline
    Category achievement type & \textit{Scalability, Time-to-solution} \\
     \hline
    Type of method used & \makecell[c]{\textit{Explicit numerical model,}\\ \textit{Dense Vision Transformer model}} \\
     \hline
    Results reported & \textit{Whole application except I/O} \\
     \hline
    Precision reported & \textit{Single precision} \\
     \hline
    System scale & \textit{Results measured on full-scale system} \\
     \hline
    Measurement mechanism & \textit{Timer, Simulated-Years-Per-Day} \\
    \hline
    \end{tabular}%
    }}

\newpage
\section{Overview of the Problem}
% description of the problem and its importance, in terms understandable to a non-specialist (1 p max)

With the legend of Yu the Great taming the floods (大禹治水) deeply rooted in Chinese history, East Asian societies have long lived with the opportunities and risks of the summer monsoon. In the summer of 2020, persistent rainfall across Yangtze River and southern China evolved into a prolonged flood disaster that affected more than 60 million people and caused direct economic losses exceeding 250 billion yuan \cite{mem2020flood,zhou2021historic}. This continuing vulnerability makes skillful flood-season forecasting operationally important for East Asia, where it directly informs flood control, water-resource management, agriculture, energy dispatch, and disaster preparedness in a region with dense populations and high exposure to hydroclimate extremes~\cite{zscheischler2018future,raymond2020understanding,hirabayashi2013global}.

\begin{figure}
\includegraphics[width=1\linewidth]{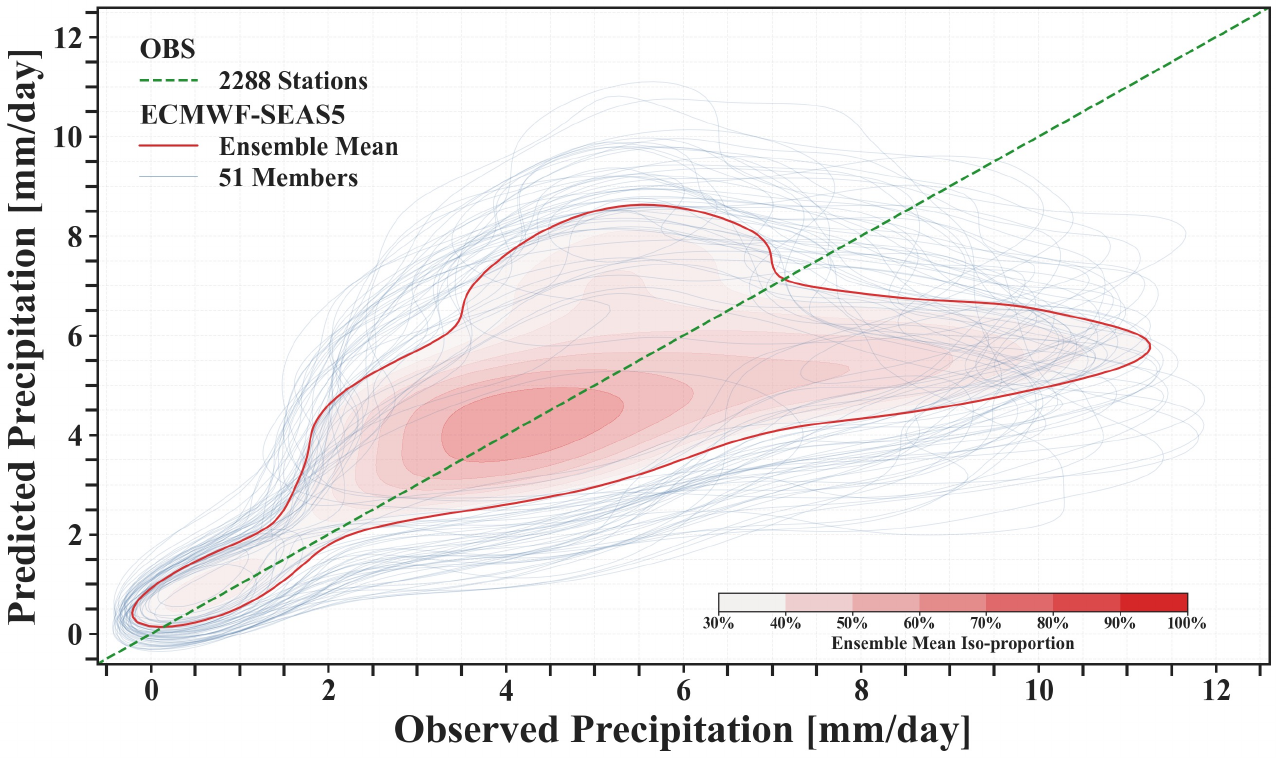}
    % \vspace{-1em}
  \caption{Comparison between observed and ECMWF-SEAS5 hindcast summer precipitation in 2020.}
  % \vspace{-2em}
  \label{fig:ec_vs_obs}
\end{figure}

In operational terms, the target problem is to predict June-to-August rainfall from March initial conditions, which is extremely difficult for several coupled reasons. First, seasonal predictability at 3- to 6-month lead times is weak because useful large-scale signals are degraded by the spring predictability barrier \cite{chen1995improved}. Second, summer rainfall over East Asia is strongly shaped by localized nonlinear convective extremes and coupled atmosphere-land-ocean interactions \cite{meehl2009decadal,koster2004regions}, which demand both physical realism and high spatial resolution. Third, operational users require not only a single forecast, but also calibrated probability distributions, regional risks, and extreme-event guidance, making large ensembles essential. Together, these requirements turn flood-season prediction into a problem that must simultaneously scale in physical fidelity, ensemble size, and computational efficiency. The difficulty in representing heavy precipitation is further illustrated by the 2020 summer hindcast of ECMWF-SEAS5 (Fig. ~\ref{fig:ec_vs_obs}). Although the ensemble reproduces the overall observed-predicted precipitation relationship, its performance deteriorates markedly for high-rainfall events. In particular, at the upper end of the observed precipitation distribution, the ensemble mean falls systematically below the one-to-one line, indicating a systematic underestimation of heavy precipitation. Meanwhile, the ensemble spread increases substantially, suggesting reduced reliability and increased uncertainty for extreme rainfall cases.

Addressing these scientific challenges first requires a high-resol\-ution coupled regional model. Such a model is needed to preserve atmosphere-land-ocean interactions, represent complex terrain effects, and retain the physical credibility required for operational seasonal prediction, especially for regional extremes. However, hig\-her resolution and tighter coupling do not solve the whole problem. As resolution increases and more small-scale processes enter the system, model complexity, parameterization burden, code maintenance, and long-term developer sustainability all become more difficult, while the cost of running large ensembles also grows rapidly with the increased resolution of the model \cite{bauer2015quiet}.

At the same time, the growing availability of denser and more diverse observations creates both opportunities and challenges. Existing numerical forecasting systems are not naturally designed to absorb and exploit all of these heterogeneous data streams efficiently across the full forecasting pipeline \cite{bauer2015quiet,brenowitz2025practical}. AI-based forecasting offers a complementary opportunity because it can learn from multi-source, multi-scale data and generate forecast samples at much lower marginal cost \cite{bi2023accurate,allen2025end,price2025probabilistic,lam2023learning,de2023machine}. However, current AI weather and climate models also come with various limitations, such as overly smooth predictions, insufficient uncertainty quantification, limited integration into operational workflows, and architectures that do not always reflect the intrinsic structure and physical constraints of atmospheric dynamics.

Based on the above considerations, in this work, we try to address the challenge of flood-season forecasting through a fused approach using both high-resolution numerical models and AI models. The key is to combine physical fidelity, richer data utilization, larger ensembles, and operational post-processing within one integrated workflow, so that increases in computing resources and data availability can be translated into measurable gains in forecasting capability. In this sense, the goal is not merely to run a larger forecasting system, but to establish a practical scaling law from exascale computing to operational flood-season prediction.

The problem addressed in this work is to build such an integrated workflow on the LineShine supercomputer, an Armv9-based next-generation system with over 23,000 nodes. Its architecture supports large-scale high-resolution numerical simulations while also providing matrix acceleration capabilities for AI workloads. We combine a high-resolution coupled regional model, a data-driven AI forecasting system, large ensembles on both tracks, and a fusion strategy that turns their joint output into operational flood-season forecasts. Through this design, we seek to show that numerical modeling, AI, data, and exascale computing can be organized into a unified forecasting framework, providing initial evidence for a practical scaling law in this challenging forecasting regime.

\section{Current State of the Art}
% quantitative discussion of current SoA, with emphasis on performance-related aspects  (1 p max)
% 是否从现有seasonal prediction的覆盖范围/网格数/分辨率/集合成员来调研SOTA？

% Operational forecasting
\subsection{Operational Seasonal Forecasting Systems}
Operational seasonal forecasting systems represent the current state of the art in ensemble-based Earth system prediction. Leading meteorological centers, including the ECMWF, UK Met Office, Météo-France, and other major operational centers, routinely deploy coupled Earth System Models for seasonal prediction, as summarized in Table \ref{tab:operational_center}.

% Table generated by Excel2LaTeX from sheet 'Sheet4'
\begin{table*}[h]
  \centering
  \caption{Operational ensemble seasonal forecasting systems from leading meteorological centers. Horizontal resolutions are approximated by their mid-latitude values for ease of comparison.}
  \vspace{-1em}
  \small
    \begin{tabular}{ccccccc}
    \toprule
    Center & NWP System & Atmosphere & Land  & Ocean & Resolution [km] & Ensemble \\
    \midrule
    ECMWF~\cite{johnson2019seas5} & SEAS5 & \multicolumn{2}{c}{IFS Cycle 43r1} & NEMO v3.4 & A36-L36-O20 & 51 \\
    UK Met Office~\cite{xavier2023assessment} & GloSea6-GC5.1 & \multicolumn{2}{c}{Met Office UM} & NEMO v4.0.4 & A60-L60-O20 & 56 \\
    JMA~\cite{Kubo2025CPS4}   & CPS4  & \multicolumn{2}{c}{JMA-GSM} & MRI.COM v5.0 & A55-L55-O20 & 145 \\
    NCEP~\cite{saha2014ncep}  & CFSv2 & \multicolumn{2}{c}{NCEP Global Forecast System} & MOM4  & A110-L110-O40 & 116 \\
    Météo-France~\cite{specq2024documentation} & System9 & ARPEGE v6.5 & SURFEX v8.0 & NEMO v4.2.0 & A50-L50-O20 & 51 \\
    DWD~\cite{doi:10.1002/essoar.10502582.1}   & GCFS2.2 & ECHAM 6.3.05 & JSBACH 3.20p1 & MPIOM 1.6.3 & A100-L100-O40 & 50 \\
    CMCC~\cite{sanna2017cmcc}  & CMCC-SPS4 & CAM6  & CLM 5.1 & NEMO v4.2  & A50-L50-O28 & 50 \\
    ECCC~\cite{Diro2024CanSIPSv3}  & CanSIPSv3.0 & GEM5.2 & ISBA  & NEMO v3.6 & A110-L110-O100 & 20 \\
    BOM~\cite{10.1071/ES22026}   & ACCESS-S2 & UM8.6   & JULES   & NEMO v3.4 & A60-L60-O20 & 121 \\
    \midrule
    Our Work   & CRESM / AI & CWRF / ERA5   & CoLM / ERA5-Land  & UOM / ORAS5 & A15-L15-O15 / A25-L25-O25 & 174 / 1,600 \\
    \bottomrule
    \end{tabular}%
    \vspace{-1em}
  \label{tab:operational_center}%
\end{table*}%

These operational systems reflect the practical trade-offs that currently shape seasonal prediction. Their horizontal resolutions of the atmosphere component are typically on the order of several tens to more than one hundred kilometers, and their ensemble sizes are constrained by the high computational cost of long-range coupled integrations. As a result, present-day operational systems are primarily optimized for robust large-scale seasonal outlooks, rather than for resolving fine-scale regional processes or supporting extremely large ensembles at high resolution.

These limitations are especially relevant for flood-season precipitation forecasting over East Asia, where skillful prediction in this setting requires both high resolution to better represent regional topography and monsoon characteristics, together with a coupled modeling framework tailored to the regional climate background. Meanwhile, further enlarging ensembles through numerical integration alone quickly becomes prohibitively expensive. Therefore, despite the maturity of current operational systems, achieving both high resolution and sufficiently large ensemble size remains a central challenge in operational seasonal forecasting.

% km-scale numerical models
\subsection{Kilometer-scale Climate Modeling}
% Table generated by Excel2LaTeX from sheet 'Sheet1'
\begin{table*}[h]
  \centering
  \caption{The comparisons among recent kilometer-resolution climate modeling efforts.}
  \vspace{-1em}
  % \scriptsize
  \resizebox{\textwidth}{!}{%
    \begin{tabular}{ccccccccc}
    \toprule
    Model & COSMO \cite{fuhrer2018near} & SCREAM \cite{taylor2023simple} & nextGEMS \cite{segura2025nextgems} & ICON-Sapphire \cite{hohenegger2023icon} & CESM-HR \cite{duan2024kilometer} & ICON~\cite{klocke2025computing}  & AP3ESM \cite{xu2025kilometer} & CRESM (ours) \\
    \midrule
    Model Components & Atmosphere & Atmosphere & Coupled & Coupled & Coupled & Coupled & Coupled & Coupled \\
    Resolution & 930 m / 1.9 km & 3.25 km & A-10 km, O-5 km & 5 km  & A-5km, O-3km & 1.25 km & 1 km  & 1 km \\
    Domain Size & Near global & global & global & global & global & global & global & Regional\times 42 Ensembles \\
    \midrule
    Scale & \makecell[c]{4,888 \\ Piz Daint GPUs} & \makecell[c]{32,768 \\ Frontier GPUs} & \makecell[c]{269 \\Levante Nodes} & \makecell[c]{600 \\Levante Nodes} & \makecell[c]{39.7M \\Sunway cores} & \makecell[c]{20,480 \\ JUPITER GPUs} & \makecell[c]{37.2M \\Sunway cores} & \makecell[c]{13.3M \\LX2 CPU cores} \\
    Integration Steps &6s / 12s&d-8.33s, p-100s&A-40s,O-80s&A-40s, O-80s&A-30s, O-90s, C-180s&A-10s, O-60s       &A-8s, O-2s       & A-9s, L-600s\\
    SDPD  & 15.7 / 83.9 & 459.9 & 600   & 126   & 222   & 145.7 & 197.1 & 27.3 \\
    \bottomrule
    \end{tabular}%
    }
  \label{tab:comparison}%
  \vspace{-1em}
\end{table*}%

Kilometer-scale Earth system modeling has emerged as a promising direction for improving regional precipitation prediction, as climate models at $O$(100 km) resolution cannot explicitly resolve key processes such as deep convection and fine-scale topographic effects \cite{palmer2014more,stevens2019dyamond}. However, moving from tens of kilometers to kilometer scale leads to a dramatic increase in computational cost, because of the enlarged mesh, reduced time step, and intensified pressure on memory, data movement, and I/O.

Early efforts mainly established the feasibility of high-resolution atmospheric modeling. For example, COSMO achieved 0.043 SYPD at 930 m resolution and 0.23 SYPD at 1.9 km on 4,888 GPUs of Piz Daint, while revealing strong memory-bound characteristics \cite{fuhrer2018near}. SCREAM further demonstrated atmosphere-only exascale capability, achieving 1.26 SYPD on 32,768 GPUs of Frontier \cite{taylor2023simple}. Beyond short capability tests, several studies have also extended high-resolution atmospheric modeling toward longer integrations. Furthermore, NICAM \cite{kodama201520} reported a 20-year global atmospheric integration at 14 km resolution with 640 cores on K computer, while ECMWF IFS \cite{wedi2020baseline} demonstrated a four-month global simulation at 1.4 km using 960 Summit nodes, with 73 and 7 SDPD, respectively.

% Table \ref{tab:comparison} summarizes the comparisons among recent kilometer-resolution climate modeling efforts. Early efforts primarily established the feasibility of kilometer-scale atmospheric modeling. For example, COSMO achieved 0.043 SYPD at 930 m resolution and 0.23 SYPD at 1.9 km on 4,888 GPUs of Piz Daint, while revealing strong memory-bound characteristics\cite{fuhrer2018near}. More recently, SCREAM demonstrated atmosphere-only exascale capability, achieving 1.26 SYPD on 32,768 GPUs of Frontier\cite{taylor2023simple}.

A further step has been the extension toward coupled Earth system configurations. The atmosphere-ocean-land coupled model ICON-Sapphire achieves 126 SDPD at 5 km resolution but only 2.5 SDPD at 1.25 km, highlighting load imbalance as a major bottleneck~\cite{hohenegger2023icon}. CESM-HR reached 222 SDPD for a coupled model with a 5-km atmosphere and a 3-km ocean on 40-million-core Sunway system \cite{duan2024kilometer}. nextGEMS further moved kilometer-scale ESM toward long-term runs, reporting about 414 and 600 SDPD for ICON and IFS-FESOM, respectively~\cite{segura2025nextgems}. More recently, AP3ESM achieved 0.54 SYPD for a 1 km AI-powered coupled Earth system model on 37.2 million Sunway cores by leveraging the performance portability features of Kokkos and OpenMP~\cite{xu2025kilometer}, while ICON reached 145.7 SDPD for a fully coupled 1.25 km global simulation on 20,480 GPUs of JUPITER~\cite{klocke2025computing}. These studies demonstrate that coupled kilometer-scale Earth system modeling is becoming technically feasible, but also show that throughput remains highly sensitive to system balance and computational scale. 

Despite these advances (see Table~\ref{tab:comparison}), existing kilometer-scale studies still fall short of large-scale, long-term ensemble forecasting. Most efforts focus either on atmosphere-only configurations or limited-duration simulations, while coupled configurations are typically constrained to capability demonstrations. As a result, it remains difficult to simultaneously achieve high resolution, long integration, and large ensemble size within a practical or operational forecasting workflow.

In this work, instead of attempting to perform large-ensemble seasonal forecasting entirely at kilometer-scale resolution, we separate the roles of resolution and throughput within a unified framework. Specifically, we optimize CRESM at 1-km resolution to establish the feasibility and performance of kilometer-scale coupled modeling, demonstrating its capability to realistically capture extreme weather processes such as typhoon evolution. At the same time, we conduct large-scale seasonal ensemble forecasting at 15-km resolution, where long integrations and thousands of ensemble members become computationally tractable. This design enables us to retain the benefits of high-resolution modeling while achieving the scale required for operational seasonal forecasting.

\section{Innovations Realized}

\begin{figure*}[h]
\includegraphics[width=1\linewidth]{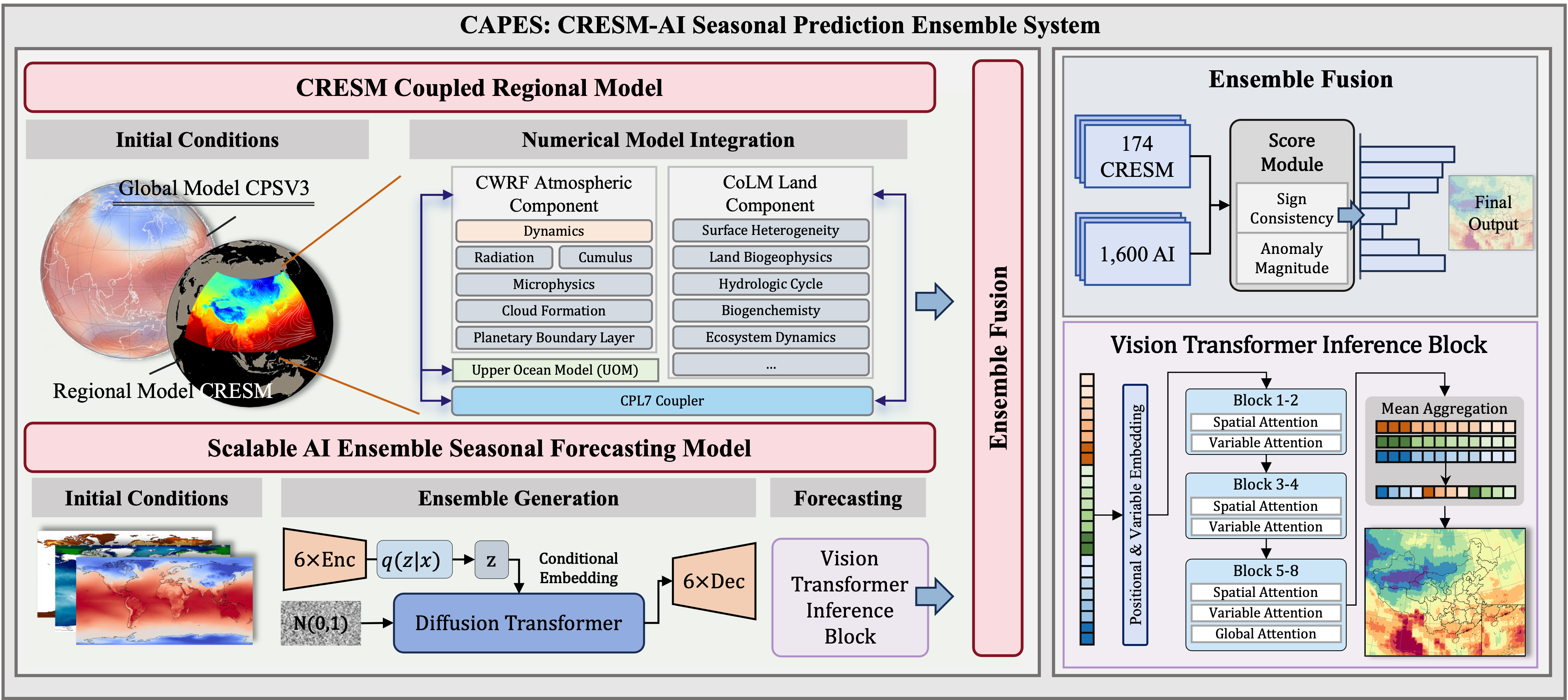}
%\vspace{-2em}
  \caption{Overview of CAPES, a hybrid forecasting workflow coupling CRESM for numerical simulation, an AI forecasting model for scalable ensemble expansion, and an ensemble-fusion module to produce a unified operational forecast product.}
  \label{fig:overall_workflow}
  \vspace{-1em}
\end{figure*}

Our innovations are realized via the \underline{C}RESM-\underline{A}I Seasonal \underline{P}rediction \underline{E}nsemble \underline{S}ystem (CAPES), a hybrid forecasting workflow that couples numerical and AI-based seasonal prediction, large-ensemble construction, and system-level co-design on the LineShine supercomputer, as shown in Fig.~\ref{fig:overall_workflow}. The numerical track provides the physically grounded regional forecasting backbone, the AI track provides scalable data-driven ensemble expansion, and the final fusion stage converts both sources into a unified operational forecast product. Together, these innovations are designed to improve flood-season forecasting capability by scaling physical fidelity, ensemble size, and computing efficiency at the same time.

\subsection{CRESM: A Coupled Regional Model}
We develop CRESM as a coupled regional Earth system forecasting framework for flood-season prediction, integrating the atmosphere, land, and ocean within one regional modeling system, as shown in Fig.~\ref{fig:overall_workflow}. Built on the CPL7-MCT coupling framework, CRESM couples the regional climate model CWRF \cite{liang2012cwrf} with the land surface model CoLM \cite{dai2003common} and a simplified upper-ocean model, enabling interactive atmosphere-land-ocean processes at regional scale.

Within CRESM, CWRF provides a nonhydrostatic atmospheric core derived from the ARW dynamical core of WRF, while CoLM contributes advanced land-surface process representation with subgrid heterogeneity. By extending the CWRF model into a coupled Earth system framework, CRESM provides a more physically complete and prediction-oriented regional modeling system, and has demonstrated improved skill in climate prediction. In this work, CRESM serves as the physical backbone for 15-km seasonal flood-season forecasting and 1-km high-resolution extreme-weather simulation.

\subsection{Optimizing CRESM for LineShine}

\subsubsection{Hybrid Parallelization Strategy}
\begin{figure}[h]
\includegraphics[width=1\linewidth]{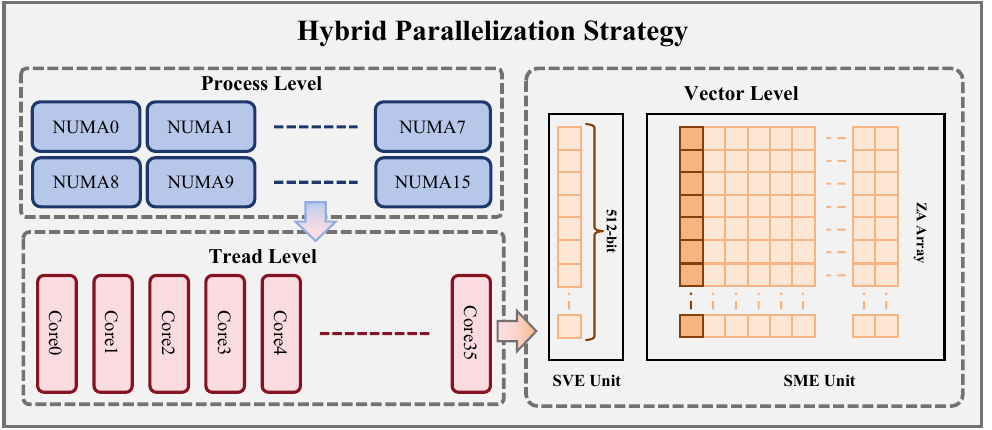}
%\vspace{-2em}
  \caption{Three-level hybrid parallelization strategy for CRESM on LineShine, with MPI domain decomposition, thread-level workload partitioning, and SIMD vectorization.}
  % \vspace{-2em}
  \label{fig:parallel_strategy}
\end{figure}

We designed a three-level hybrid parallelization strategy spanning MPI, threading, and SIMD to fully exploit the massive cores and Scalable Vector Extension (SVE) units of the processor, as shown in Fig.~\ref{fig:parallel_strategy}.

At the process level, MPI is used for domain decomposition, partitioning the grids into subdomains along \(i\) and \(j\) directions. Each process performs local computation and maintains halo regions for inter-process communication.

At the thread level, we employ two complementary strategies to improve shared-memory utilization. A grid-based decomposition partitions each subdomain across threads for kernels without horizontal data dependencies. This approach incurs low runtime overhead and is well suited for physics kernels without horizontal data dependencies, but cannot guarantee balanced computational workloads across threads. We complement it with direct task partitioning over loops inside individual computational kernels. Compared with grid-based decomposition, this method is more flexible and is particularly suitable for dynamical-core kernels with data dependencies in both the \(i\) and \(j\) directions. To control its runtime overhead, we replace OpenMP with a vendor-provided pthread library, reducing overhead by more than 50\% while enabling dynamic task stealing among threads for better load balance.

At the vectorized level, we further adapt CRESM to exploit the 512-bit SVE units of the processor through SIMD optimization. Combining compiler auto-vectorization with vectorized math libraries, we successfully vectorize most transcendental operations, including \texttt{expf}, \texttt{powf}, and \texttt{logf}. For the large number of stencil computations, we employ a pattern recognition-based method to split and restructure specific loop kernels, and implement SVE and SME intrinsics in C codes. Because loop-level task partitioning is introduced at the thread-parallel level, data blocks can be aligned more naturally with the vector width of the hardware, improving vector-unit utilization and overall SIMD efficiency.

\subsubsection{Memory and I/O Optimization}
At high resolution, CRESM exhibits two major memory-related bottlenecks: inefficient memory access caused by its fixed global data layout, and a substantial memory footprint. This is particularly important on the target LineShine platform, where each NUMA node is equipped with High Bandwidth Memory (HBM). Although HBM provides high memory bandwidth to mitigate memory-access bottlenecks, its limited capacity requires both bandwidth-efficient data access and careful control of memory usage.

\begin{figure}
\includegraphics[width=1\linewidth]{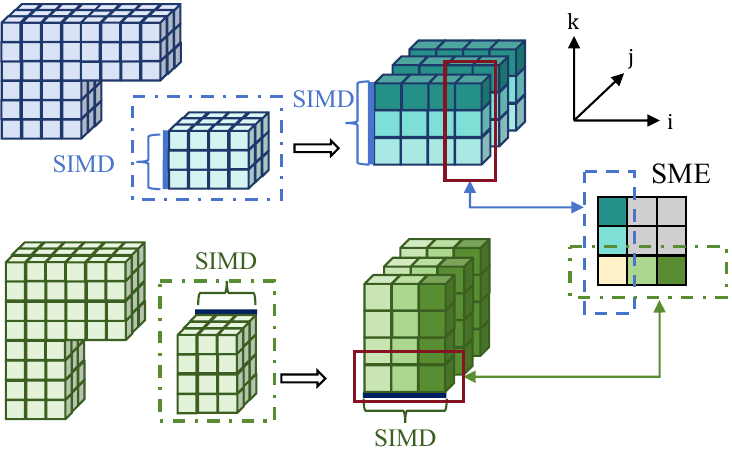}
\vspace{-3em}
  \caption{Memory-layout transformation for CRESM on LineShine.}
  \label{fig:memory}
  \vspace{-2em}
\end{figure}

To address the memory-access inefficiency, we introduce a global memory layout transformation interface, as illustrated in Fig.~\ref{fig:memory}. This interface utilizes multi-core parallel SME units to perform low-level layout conversions on the data flow between the dynamical core and physical parameterization kernels. Most physics kernels operate on single-column data, where data dependencies exist only along the vertical \(k\) dimension. In contrast, the dynamical core involves advection computations with dependencies in the horizontal \(i\) and \(j\) dimensions. Through layout transformation, data can be arranged contiguously along dependency-free dimensions while the tile size is aligned with the SIMD vector width, improving locality and enabling more effective utilization of HBM bandwidth.

To reduce the memory footprint of CRESM, we apply several optimization strategies. First, we reduce the number of MPI processes while increasing thread parallelism, thereby lowering the extra halo storage overhead associated with process-level domain decomposition. Second, we refactor the original single-process I/O implementation in CWRF into a fully distributed I/O model, allowing each process to perform I/O independently. This eliminates the large buffers previously required by dedicated I/O processes and prevents severe memory imbalance across processes. For global data access in CWRF, we further reconstruct all read and write operations into an on-demand pattern based on spatial and temporal requirements, minimizing the size of intermediate data buffers. For the grouped I/O implementation in CoLM, we introduce a finer-grained grid decomposition during the data preprocessing stage and employ more I/O process groups to amortize buffer overhead. Finally, we identify the input and output variables within functions to detect high-dimensional temporary arrays propagated across loops. By fusing adjacent loops, we avoid passing temporary variables through high-dimensional intermediate arrays, which significantly reduces runtime memory overhead.

\subsubsection{Communication Optimization}
Each CWRF process has eight neighboring processes corresponding to the east, west, south, north, southeast, southwest, northwest, and northeast directions. Using \texttt{RSL\_LITE} as the underlying communication library for halo exchange across \(i\)-\(j\) domain partitions, CWRF employs a two-step communication scheme for diagonal neighbors. Specifically, messages are first sent to the neighboring process in the \(j\)-direction and then forwarded in the \(i\)-direction to the target process. By utilizing the dual-plane, multi-rail, and four-level fat-tree network topology of the LineShine cluster, we implement direct point-to-point communication between diagonally adjacent processes to reduce communication latency and alleviate network bandwidth pressure.

\begin{figure}
\includegraphics[width=1\linewidth]{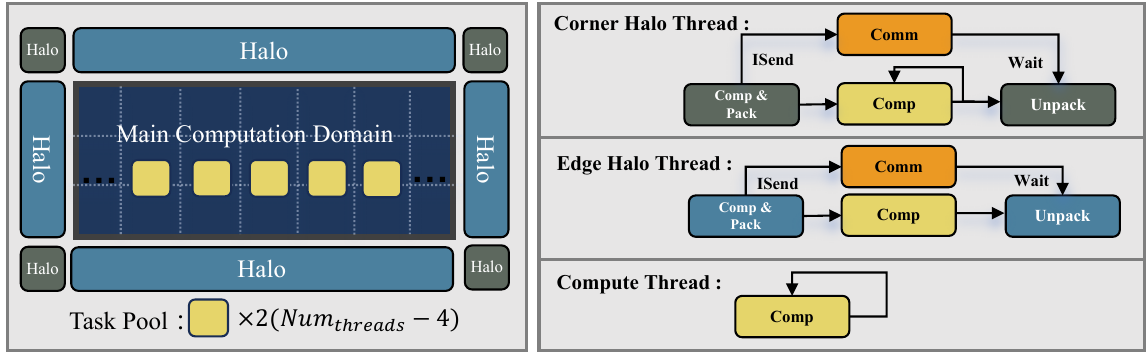}
\vspace{-2em}
  \caption{Computation-communication overlap in CWRF.}
  \label{fig:communication}
  \vspace{-2em}
\end{figure}

Furthermore, we introduce a computation-communication overlap to improve end-to-end performance. Specifically, the interior computational region of each process is partitioned into a number of blocks greater than the total number of threads, forming a task queue from which each thread repeatedly fetches work. In parallel, eight threads are assigned to handle the packing, unpacking, and asynchronous communication requests for the halo regions in the eight directions, as shown in Fig.~\ref{fig:communication}. While waiting for the asynchronous communication to complete, these threads return to the task queue and participate in interior-region computation. In this way, communication latency is effectively hidden, and thread-level load balance is improved through dynamic task scheduling.

\subsection{Scalable AI Ensemble Seasonal Forecasting}
%Model
The AI track is designed as a seasonal forecasting system that absorbs multi-source, multi-scale atmosphere-land-ocean information and expands the forecast ensemble at low marginal cost. Its inputs are derived from ERA5, ORAS5, and ERA5-Land reanalysis data, and combine day-scale, month-scale, and year-scale information so that the model can represent short-term weather states, seasonal background circulation, and low-frequency climate memory within one unified input space.

Specifically, the atmospheric profiles incorporate the past 14 days, 3 months, and 2 years of global information, while the oceanic and land-surface inputs encode seasonal and interannual memory over the same macroscopic periods. All operational forecasts are initialized on March 1st of each target year to predict the June, July, and August mean precipitation fields. This multimodal data fusion yields high-dimensional tensors spanning the atmosphere, land, and ocean, and provides the AI system with a coupled representation of recent weather evolution, climate background state, and multi-sphere interactions.

\subsubsection{Generative Ensemble Construction}
To achieve a forecasting scale of 1,600 members per year without the prohibitive computational cost of traditional dynamical data assimilation, we design a data-driven perturbation paradigm. This architecture mirrors the initial-condition uncertainty of numerical ensemble forecasting.

As shown in Fig.~\ref{fig:overall_workflow}, to generate diverse and physically credible initial atmospheric states, we deploy a multi-stream Variational Autoencoder (VAE) \cite{kingma2013auto} coupled with a Diffusion Transformer (DiT) \cite{peebles2023scalable}. The daily atmospheric inputs are decoupled into six distinct streams, with the VAE encoder independently projecting each stream into a low-dimensional latent space. After sampling, the resulting latent tensors are concatenated and passed to the DiT. By drawing independent samples from the learned diffusion process, the DiT explores the chaotic attractor of the atmospheric state, which the VAE decoder reconstructs back into physical fields and efficiently generates 40 independent initial perturbations for each year.

To represent forecast-time uncertainty, we introduce a second perturbation within the forecasting backbone. During inference, targeted stochastic noise is injected directly into the deep latent spaces of the forecasting model. By applying 40 unique latent perturbations to each of the 40 generated initial conditions, the system produces 1,600 distinct AI ensemble members per year. This dual-perturbation strategy provides a large probabilistic envelope while remaining entirely compute-bound and highly parallelizable.

\subsubsection{Forecasting Backbone}
To process the massive multi-domain data, the atmospheric, oceanic, and land-surface variables are first compressed to 16 channels via Principal Component Analysis (PCA). Rather than stacking these heterogeneous domains along the channel dimension, which would prematurely entangle distinct physical processes of each Earth system component during the early stage, the compressed fields are patched and concatenated along the sequence dimension, yielding an extreme sequence length of 777,600 tokens in total.

\begin{figure}
\includegraphics[width=1\linewidth]{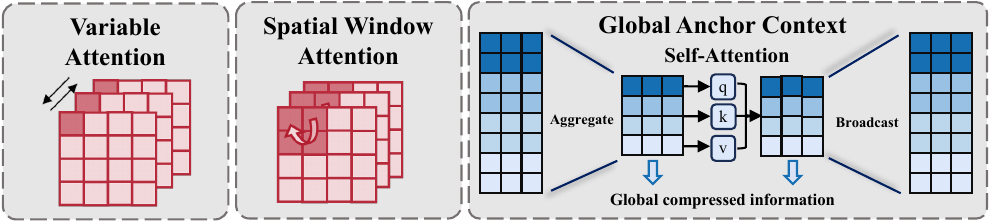}
\vspace{-2em}
  \caption{Customized tri-level attention architecture for efficient Earth system forecasting.}
  \label{fig:ViT_attention}
  \vspace{-2em}
\end{figure}

A standard self-attention mechanism would incur a quadratic $O(L^2)$ computational and memory cost, rendering inference on this scale intractable. To overcome this bottleneck, our forecasting backbone integrates a customized tri-level attention architecture, as shown in Fig.~\ref{fig:ViT_attention}. Window-based spatial attention captures localized fluid dynamics and mesoscale structures within bounded regional grids. Cross-variable attention synchronizes thermodynamic dependencies among the tightly coupled atmospheric, oceanic, and land-surface channels at each spatial location. A global anchor context employs a compressed set of latent anchors to aggregate and broadcast information across the globe, preserving large-scale teleconnections essential for seasonal precipitation prediction without computing a dense $O(L^2)$ attention matrix. This design reduces the overall attention complexity from quadratic $O(L^2)$ to about $O(L)$, slashing the memory footprint and enabling high-throughput inference. Finally, a lightweight decoder projects the representations back into monthly precipitation fields at native spatial resolution.

\subsection{Optimizing the AI Module for LineShine}

\subsubsection{Memory-Efficient Long-Sequence Inference}
The AI seasonal forecasting model must sustain long-sequence inference over coupled atmosphere-land-ocean inputs while also serving as the large-scale ensemble expansion engine of the overall workflow. The tri-level attention design described above therefore functions not only as a modeling innovation, but also as a throughput optimization that makes CPU-based seasonal ensemble production feasible.

\subsubsection{Hardware Adaptation to the LineShine Platform}
Hardware adaptation of the AI module to the LineShine platform is performed at the operator level to exploit the SME/SVE matrix and vector acceleration capabilities of the Armv9-based many-core architecture. For compute-intensive linear layers, we integrate a GEMM operator optimized for SME/SVE, featuring runtime tiling and workload partitioning to improve cache locality and multi-core utilization. 
%For memory-sensitive non-linear operators, including Layer Normalization and activation functions, we develop SVE-vectorized implementations to reduce data movement and improve throughput. These optimizations enable efficient Transformer inference on the CPU-based LineShine system.
For memory-sensitive non-linear operators such as Layer Normalization and activation functions, we develop SVE-vectorized implementations to reduce data movement and improve throughput, enabling efficient Transformer inference on the LineShine system.

\subsection{Hybrid Ensemble Design, Fusion, and System Integration}
The final innovation is the construction of a large hybrid ensemble that combines the complementary strengths of the numerical and AI tracks within one operational workflow. On the numerical side, we build a 174-member ensemble in two stages. The first 27 members are generated from three start dates and nine physics-scheme combinations inherited from the existing seasonal forecasting configuration. The remaining 147 members are generated from the same three start dates together with a $7 \times 7$ sweep over two key physical parameters around a strong-performing parameterization setting. This design preserves both structural diversity across physics schemes and controlled perturbation of sensitive physical parameters.

On the AI side, we construct a 1,600-member ensemble by combining 40 generative initial perturbations with 40 perturbations applied within the forecasting model. These two stages mirror, in data-driven form, the roles of initial-condition perturbations and model perturbations in traditional ensemble forecasting. The AI ensemble therefore serves as a low-cost but large-scale expansion of forecast uncertainty, while the numerical ensemble remains the physically grounded backbone of the system.

The outputs of both tracks are combined through an adaptive fusion and post-processing stage to form the final 1,774-member probabilistic forecast. The fusion module quantifies the contribution of each ensemble member using two complementary metrics. The first measures sign consistency by comparing each member's precipitation anomaly with the ensemble-median anomaly, reflecting its robustness in capturing the dominant signal. The second measures anomaly magnitude relative to climatology, reflecting the member's sensitivity to anomalous conditions such as droughts or excessive rainfall. These two metrics characterize robustness and anomaly sensitivity, respectively, and are normalized and combined into a unified contribution score for each member. Within the large-ensemble framework, this strategy exploits inter-member diversity by jointly emphasizing agreement with the dominant signal and responsiveness to anomaly amplitude, leading to a more realistic representation of precipitation anomalies and improved seasonal prediction skill.

At the system level, the numerical and AI workloads are mapped concurrently onto the LineShine supercomputer so that both tracks can execute within one operational workflow. This design transforms CAPES from a collection of separate models into a unified forecasting system in which coupled simulation, AI prediction, ensemble construction, and fusion jointly contribute to operational flood-season forecasting capability.

\section{How Performance was Measured}
% (Note that preference is given to performance actually measured [not projected], based on the entire application [including I/O] and with uniform precision.  Explain in detail if any portion of total runtime was not included in the measurements, if and where different precisions were used, or any attributes listed in Section 3 as “other”).

% what application(s) was used to measure performance (1 p max)
\subsection{CAPES Configuration}
Table \ref{tab:model_configuration} summarizes the six CRESM configurations used in the performance evaluation, covering horizontal resolutions from 1 km to 30 km. For all configurations, the numbers of vertical layers in CWRF, CoLM, and UOM are 36, 10 and 30, respectively. 

% Table generated by Excel2LaTeX from sheet 'Sheet1'
\begin{table}[h]
  \centering
  \caption{Major configurations of CRESM resolution used in measurement.}
  \vspace{-1em}
  \resizebox{\linewidth}{!}{%
    \begin{tabular}{ccrcrcr}
    \toprule
          & \multicolumn{2}{c}{CWRF (ATM)} & \multicolumn{2}{c}{CoLM (LAND)} & \multicolumn{2}{c}{UOM (OCEAN)} \\
    \midrule
    \multicolumn{1}{l}{Res. (km)} & \multicolumn{1}{c}{Int. T [s]} & \multicolumn{1}{l}{Grids No.} & \multicolumn{1}{l}{Int. T [s]} & \multicolumn{1}{l}{Grids No.} & \multicolumn{1}{l}{Int. T [s]} & \multicolumn{1}{l}{Grids No.} \\
    \midrule
    1     & 9 (Topo smoothed)    & 35,550,900 & 600  & 22,317,046 & 120  & 13,848,008  \\
    3     & 10   & 3,950,100 & 600  & 2,712,658 & 120  & 1,526,810  \\
    6     & 20   & 987,525 & 600  & 758,954 & 120  & 378,680  \\
    10    & 50   & 355,509 & 600  & 311,243 & 120  & 134,792  \\
    15    & 60   & 158,004 & 600  & 158,515 & 120  & 59,114  \\
    30    & 120  & 39,501 & 600  & 52,918 & 120  & 14,245  \\
    \bottomrule
    \end{tabular}%
    }
    \vspace{-1em}
  \label{tab:model_configuration}%
\end{table}%

For the AI component, the generative model contains 2.14 B parameters, including six VAEs with 0.35 B parameters each and one DiT-S/2 with 34.3 M parameters. The ViT forecasting backbone contains 0.6 B parameters, with an embedding dimension of 512, 8 Transformer layers, 8 attention heads, and a patch size of 8.

\subsection{Performance Metrics}
For our experiments, we evaluate both computational performance and forecasting skill. Computational performance is measured using the time-to-solution of the full operational forecasting workflow, excluding I/O, over a 10-year hindcast period. We also report the strong-scaling and weak-scaling efficiency of CRESM to quantify its parallel scalability on the LineShine system. For numerical simulations of CRESM, performance is also expressed in simulated years per day (SYPD) and simulated days per day (SDPD), computed as the ratio between the length of simulated time and the corresponding wall-clock execution time. 
% Unless otherwise noted, the SYPD metric excludes initialization and I/O overheads.

% , whereas the end-to-end time-to-solution \hl{includes} the complete operational workflow.

Forecast skill is evaluated using the PS Score, an operational metric utilized by the China Meteorological Administration (CMA) for seasonal precipitation forecasting. Specifically, rainfall anomalies are classified into normal, first-level anomaly ($N1$, precipitation anomaly percentage in the ranges 20\%$\sim$50\% or -50\%$\sim$-20\%), and second-level anomaly ($N2$, >50\% or <-50\%) categories. In addition, a penalty term $M$ is assigned when the observed precipitation is extreme (>100\% or <-100\%) but the forecast fails to reach $\pm$50\%. Let $N0$ denote the number of samples with the correct anomaly sign, the PS is computed in Eq.~\ref{eq:PS_score}. Compared with traditional error metrics such as RMSE, PS better reflects the requirements of operational seasonal forecasting, where correctly identifying anomalous and extreme precipitation is critical for decision-making in flood control and water resource management. 
\begin{equation}
    PS = \frac{2\times N_0 + 2\times N_1 + 4\times N_2}{(N-N_0) + 2\times N_0 + 2\times N_1 + 4\times N_2 + M}\times 100
    \label{eq:PS_score}
\end{equation}

% system and environment where performance was measured (1 p max)
\subsection{System Details}
\begin{figure}
\includegraphics[width=1\linewidth]{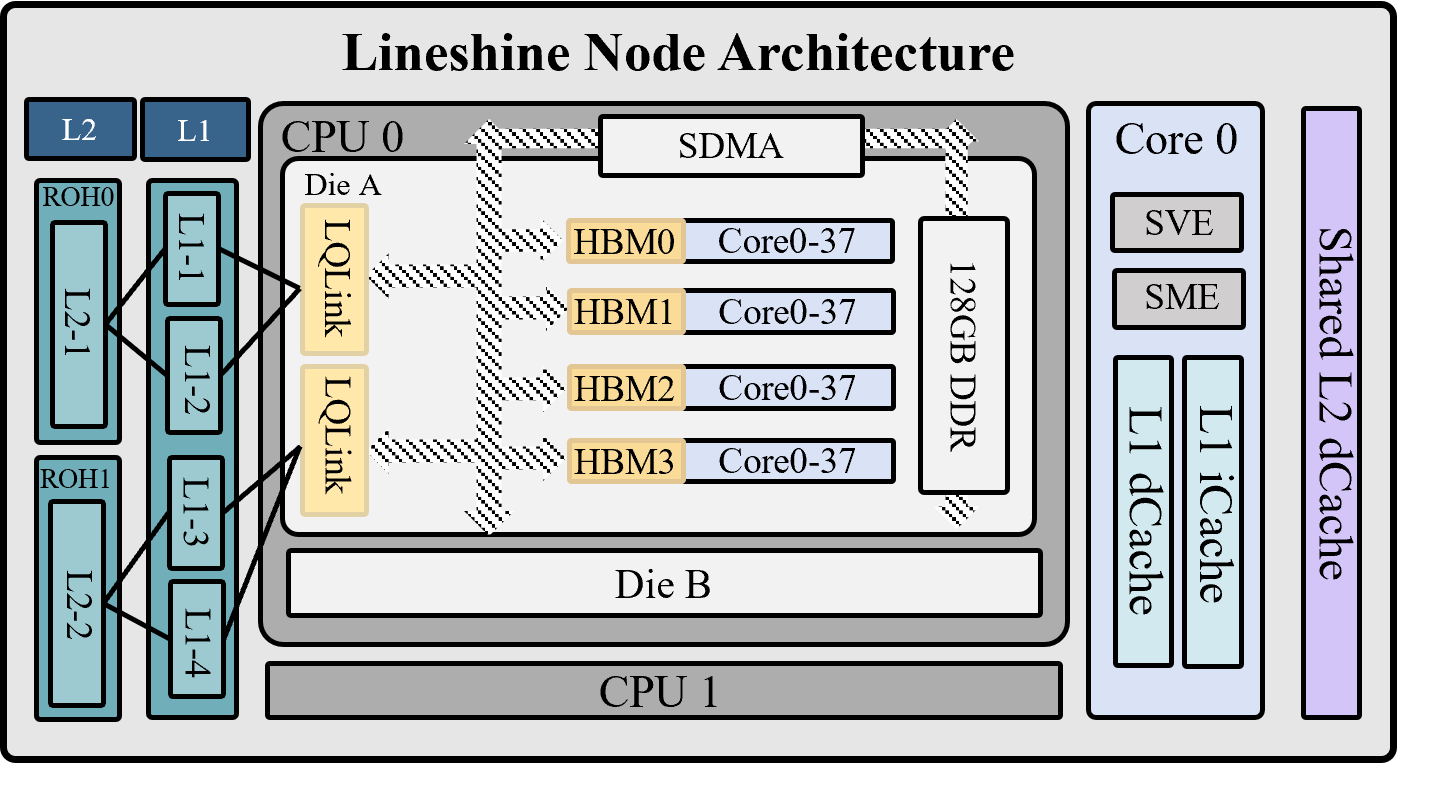}
  \vspace{-2em}
  \caption{The LineShine supercomputer structure.}
  \label{fig:supercomputer_structure}
  \vspace{-1em}
\end{figure}

Our performance evaluations are conducted on the LineShine supercomputer deployed at the National Supercomputing Center in Shenzhen, China, as shown in Fig. ~\ref{fig:supercomputer_structure}. The system contains more than 23,000 compute nodes. Each node has two Armv9-based LX2 processors. Each processor integrates two compute dies and provides 304 cores in total, giving 608 cores per node. Each LX2 processor is equipped with eight on-package HBM stacks, providing 32 GB capacity and 4 TB/s aggregate bandwidth. In addition, each compute die is paired with 128 GB DDR memory organized into four NUMA domains, resulting in 256 GB DDR per processor. Each die also includes an SDMA engine for data transfer between DDR and HBM.

The LX2 processor supports FP64, FP32, FP16, and INT8 arithmetic through SME and SVE units, with a peak FP64 performance of up to 60.3 TFLOP/s. All nodes are interconnected by the LingQi network, which uses a dual-plane, multi-rail fat-tree topology and provides 1.6 Tb/s bandwidth per node. In our evaluation, we employed two configurations: one process per NUMA node with 36 threads per process, and four processes per NUMA node with 9 threads per process.

\section{Performance Results}
% include scalability (weak and strong), time to solution, efficiency (of bottleneck resources), and peak performance (2 pp max)

\subsection{Strong Scalability}
\begin{figure*}
\includegraphics[width=1\linewidth]{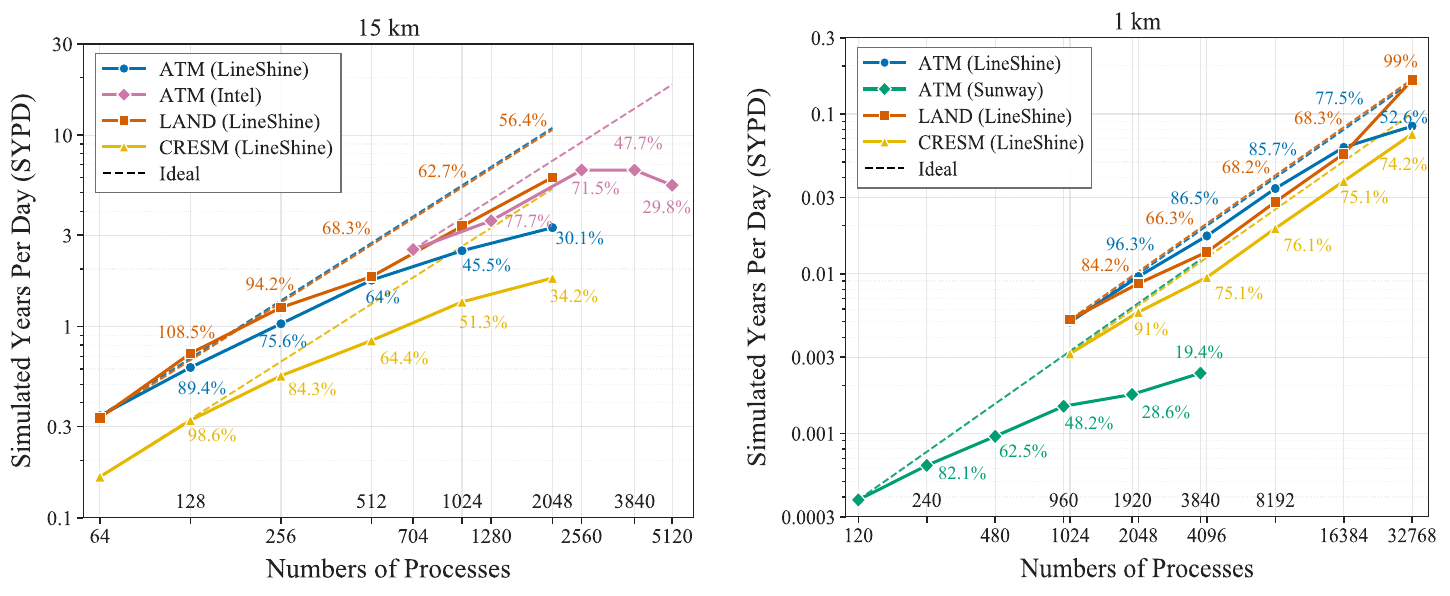}
% \vspace{-1em}
  \caption{Strong scalability of the atmosphere model (ATM), the land model (LAND) and the CRESM coupled model in 15 km and 1 km resolutions compared with Intel-6458Q and Sunway platform.}
  \label{fig:performance_strong_scaling}
  \vspace{-1em}
\end{figure*}

The strong-scaling results are shown in Fig.~\ref{fig:performance_strong_scaling} for the 15-km and 1-km configurations of CRESM, together with the corresponding atmospheric (ATM) and land (LAND) components. For comparison, we also evaluate the CWRF atmospheric component on an Intel-6458Q-based platform at 15-km resolution and on the Sunway OceanLight system at 1-km resolution, in order to assess the effectiveness of the LineShine port and optimization.

We first examine the atmospheric component. On LineShine, the 15-km CWRF configuration scales from 64 to 2,048 processes, with performance increasing from 0.34 SYPD to 3.28 SYPD, corresponding to a parallel efficiency of 30.1\%. For the 1-km configuration, scaling from 1,024 to 32,768 processes improves performance from 0.005 SYPD to 0.083 SYPD, achieving a higher parallel efficiency of 52.6\%. For cross-platform comparison, the 15-km CWRF configuration on the Intel platform scales from 704 to 5,120 processes with a parallel efficiency of 29.8\%, while the 1-km configuration on Sunway OceanLight scales from 120 to 3,840 processes with a parallel efficiency of 19.4\%. These results demonstrate that the strong scalability of optimized CWRF on LineShine is comparable to other architectures-adapted versions.

For the land component on LineShine, the 15-km and 1-km configurations achieve parallel efficiencies of 56.4\% and 99\%, respectively. At the coupled-model level, the full CRESM system achieves strong-scaling efficiencies of 34.2\% at 15 km and 74.2\% at 1 km. These results indicate that the optimized coupled model preserves good scalability across resolutions, and they establish the practical feasibility of both production-scale seasonal forecasting at 15 km and kilometer-scale high-resolution simulation at 1 km. This level of scalability enables a full seasonal integration at 15-km resolution to finish within 15 hours using 8 nodes, while making kilometer-scale coupled simulation computationally attainable, with the 1-km experiment completed in 7 days with 512 nodes. 

\begin{figure}
\includegraphics[width=1\linewidth]{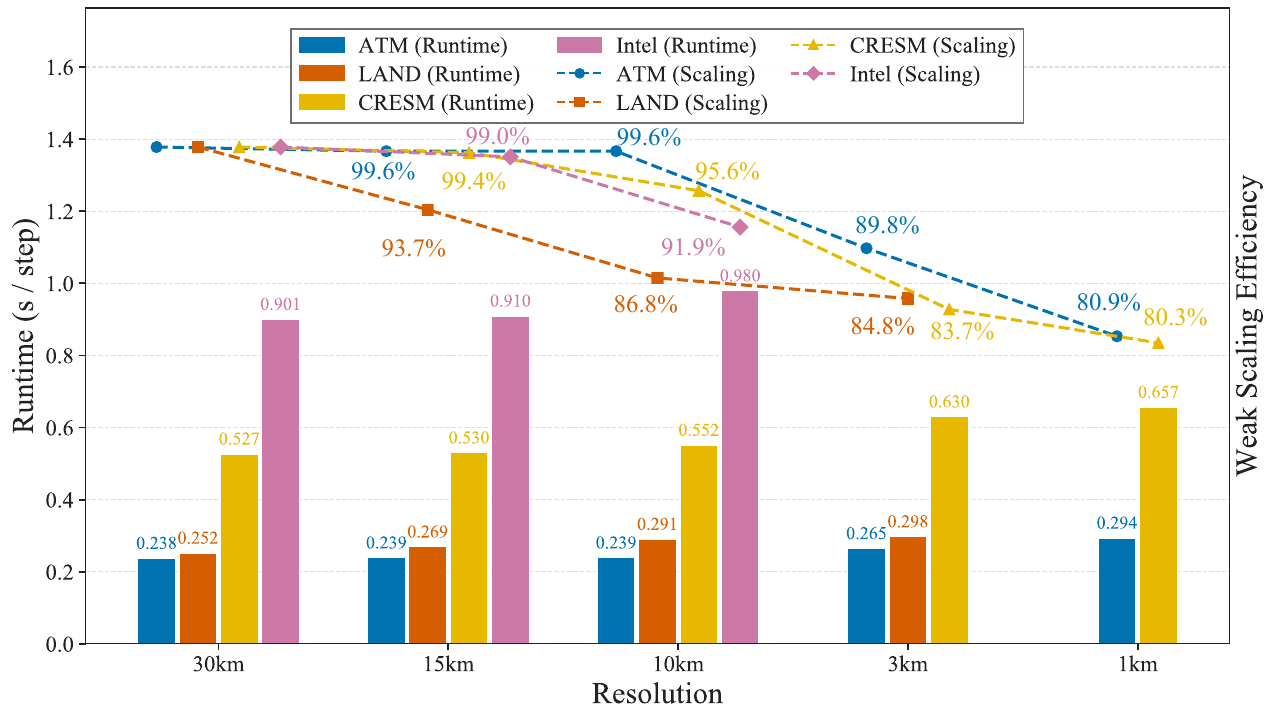}
% \vspace{-2em}
  \caption{Weak scalability of the atmosphere model, the land model, and the CRESM model, from 30 km to 1 km, compared with the Intel platform.}
  \label{fig:performance_weak_scaling}
  \vspace{-1em}
\end{figure}

\subsection{Weak Scalability}
Fig.~\ref{fig:performance_weak_scaling} shows the weak-scaling results of CRESM across the horizontal resolutions listed in Table~\ref{tab:model_configuration}. In this experiment, the problem size is increased proportionally with the number of nodes so that the workload per node remains approximately constant. We evaluate the atmosphere (ATM), land (LAND), and fully coupled CRESM configurations across six resolution settings using 1 km, 3 km, 6 km, 10 km, 15 km, 30 km.

For atmospheric component, the weak-scaling efficiency reaches 80.9\%, while the land component achieves 84.8\%. At the coupled-model level, CRESM sustains a weak-scaling efficiency of 80.3\%. This confirms the model preserves good scalability across increasing resolutions and machine scales. At 1-km resolution, the coupled model still maintains high weak-scaling efficiency, providing additional evidence that the communication and memory optimizations are effective at large scale. The remaining efficiency loss at the highest resolutions is primarily caused by increased communication latency and synchronization overhead across a larger number of nodes.

\subsection{Analysis of CRESM Optimization}

Based on the data presented in Table~\ref{tab:operator_optimization} and the optimization strategies detailed in Section 5.2, the main operators of the CRESM model achieve substantial performance gains on the LineShine platform.

For compute-intensive Microphysics and Radiation drivers, hybrid parallelization and memory layout optimization yield remarkable benefits. The introduction of thread-level parallelism delivers initial speedups of 3.72\times and 4.08\times. Subsequently, SIMD vectorization leveraging SVE/SME instructions further elevates these speedups to 5.94\times and 6.46\times. Ultimately, memory layout transformations tailored for HBM substantially enhance data spatial locality, culminating in final speedups of 8.52\times and 6.80\times for the microphysics and radiation drivers. These optimizations not only reduce execution time but also lead to a substantial reduction in Level 1 data cache misses by factors of about 3.45\times and 4.80\times. This corroborates the profound improvements in memory access efficiency.

For communication-intensive operations, the topology-aware exchange and computation-communication overlapping optimizations effectively reduce communication latency. The execution time of the halo exchange operator halo\_em\_phys\_di and halo\_em\_d2\_5 is accelerated by factors of 3.37 and 2.62, respectively.

Overall, the step-wise optimization strategy mitigates communication latency and maximizes the throughput of multi-core vector units, thereby fully unlocking the hardware potential of the LineShine cluster.

\begin{table*}
  \centering
  \caption{Step-wise optimization of main operators.}
  % \vspace{-1em}
  %\resizebox{\linewidth}{!}{%
  \begin{tabular}{lcccccc}
    \toprule
    \multirow{2}{*}{Optimization} & \multicolumn{2}{c}{Microphysics Driver} & \multicolumn{2}{c}{Radiation Driver} & \multicolumn{1}{c}{halo\_em\_phys\_di} & \multicolumn{1}{c}{halo\_em\_d2\_5} \\
    \cmidrule(lr){2-3} \cmidrule(lr){4-5} \cmidrule(lr){6-6} \cmidrule(lr){7-7}
          & Time / Speedup & L1d-miss & Time / Speedup & L1d-miss & Time / Speedup & Time / Speedup \\
    \midrule
    Baseline       & 3.23 s         & 72,563,081 & 2.59 s         & 68,760,202 & 0.236 s        & 0.854 s        \\
    + OMP          & 3.72$\times$   & --         & 4.08$\times$   & --         & --             & --             \\
    + SIMD         & 5.94$\times$   & --         & 6.46$\times$   & --         & --             & --             \\
    + HBM          & 8.52$\times$   & --         & 6.80$\times$   & --         & --             & --             \\
    + Comm opt & --             & --         & --             & --         & 0.07 s         & 0.326 s        \\
    \midrule
    Final          & 8.52$\times$  & 20,991,399 &  6.80$\times$  & 14,317,097 & 3.37$\times$   & 2.62$\times$   \\
    \bottomrule
  \end{tabular}%
  %}
  \vspace{-1em}
  \label{tab:operator_optimization}
\end{table*}

\subsection{1-km CRESM simulations}
While the 15-km configuration targets seasonal prediction of summer precipitation, the 1-km configuration extends the system toward fine-scale forecasting of extreme weather.

For the 1-km CRESM configuration, we conduct a high-resolution capability test using Super Typhoon Saola, which formed over the western North Pacific east of the Philippines on 24 August 2023. The model is initialized at 0000 UTC 31 August 2023 and integrated for 72 hours to assess its ability to reproduce the evolution and fine-scale structure of Saola. Unlike the 15-km seasonal integrations, this 1-km experiment evaluates the feasibility and performance of kilometer-scale coupled simulation, as well as its capacity to represent extreme weather processes at high spatial resolution.

\begin{figure*}[h]
\includegraphics[width=1.0\textwidth]{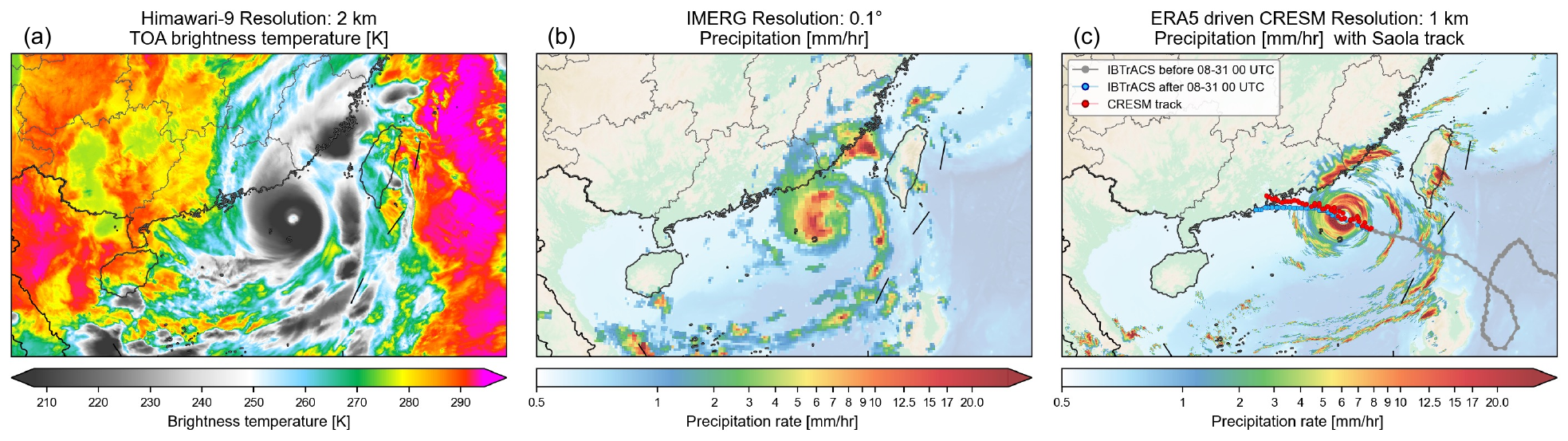}
% \vspace{-2em}
  \caption{Structures of Typhoon Saola at 1400–1500 UTC 31 August 2023 from Himawari-9 satellite observations, IMERG precipitation, and the ERA5-driven CRESM simulation. (a) Himawari-9 top-of-atmosphere brightness temperature at 2 km resolution. (b) IMERG mean precipitation at 0.1 deg resolution. (c) ERA5-driven CRESM mean precipitation at 1 km resolution with the Saola tracks from IBTrACS and CRESM.}
  \label{fig:performance_1km}
  \vspace{-0.5em}
\end{figure*}

As shown in Fig.~\ref{fig:performance_1km} (a) and (b), at 1400 UTC 31 August 2023, the Himawari-9 image shows a compact, well-defined eye embedded within a highly organized cloud shield, indicative of a mature and intense tropical cyclone with pronounced azimuthal symmetry in cloud-top structure. The IMERG precipitation field also captures the storm-scale precipitation pattern and the major spiral rainbands surrounding the vortex, with the heaviest rainfall concentrated near the inner core. However, owing to its relatively coarse spatial resolution, IMERG does not resolve the fine-scale eyewall structure and lacks a clear eye signature comparable to that seen in the satellite brightness temperature field. In addition, its broader precipitation footprint is mainly associated with weak rainfall. In our 1-km simulation shown in Fig.~\ref{fig:performance_1km} (c), the ERA5-driven CRESM reproduces the compact vortex core and spiral rainband structure of Saola, with intense precipitation concentrated near the eyewall and extending outward along curved bands. Compared with IMERG, the simulated precipitation field is more tightly organized around the storm center, while the observed rainfall appears broader and more asymmetric due to the larger area of weak precipitation in the outer region. The simulated typhoon-center trajectory also closely follows the IBTrACS track during the analyzed period, indicating that the 1-km configuration reproduces both the spatial organization of precipitation and the propagation of Saola with reasonable fidelity.

\subsection{Performance Evaluation of ViT}
We further evaluate the effectiveness of the proposed customized ViT model \texttt{ViT+PCA+Seq.}, which organizes variables along the sequence dimension, by comparing it with \texttt{ViT+PCA}, where variables are organized in the channel dimension, and with a standard ViT baseline. Specifically, \texttt{ViT+PCA} adopts the same PCA-based input dimensionality reduction as our method, enabling a controlled comparison under identical input representations and thereby isolating the effectiveness of the proposed sequence-wise formulation and tailored attention mechanisms. To ensure a fair comparison, all baseline models are configured to have a total number of parameters comparable to that of our model.

According to Table ~\ref{tab:ViT_performance}, \texttt{ViT+PCA+Seq.} achieves the best performance across all metrics, outperforming \texttt{ViT+PCA} and standard ViT. These results suggest that the gains stem from the combination of sequence-wise variable organization and tailored attention mechanisms. Compared to \texttt{ViT+PCA}, our method organizes variables along the sequence dimension, allowing variable-wise tokens to directly participate in attention instead of being mixed in channel space. This leads to more effective modeling of spatial, inter-variable, and global dependencies.

\vspace{1em}
\begin{table}
  \centering
  \caption{Performance evaluation of sequence-wise formulation and attention mechanisms in the forecasting model.}
  % \vspace{-2em}
    \begin{tabular}{cccc}
    \toprule
          & \multicolumn{1}{l}{PS$\uparrow$} & \multicolumn{1}{l}{RMSE$\downarrow$} & \multicolumn{1}{l}{ACC\_ano$\uparrow$} \\
    \midrule
    \texttt{ViT}   & 69.37 & 2.07  & 0.006 \\
    \texttt{ViT+PCA} & 65.72 & 2.21  & -0.063 \\
    \texttt{ViT+PCA+Seq.} & 72.43 & 2.02 & 0.077 \\
    \bottomrule
    \end{tabular}%
  \label{tab:ViT_performance}%
  % \vspace{-1em}
\end{table}%

\subsection{Ten-year hindcasts of CAPES}
We conduct 10-year hindcast experiments over 2016-2025 to assess CAPES. We compare its performance against hindcasts from major operational centers. Although these operational systems are generally issued with monthly initializations, the present comparison is restricted to forecasts initialized in March and integrated through the flood season. Operational centers with fewer than five valid hindcast years during 2016-2025 are excluded. For centers without complete 10-year archives, only the available valid years are included.

\begin{figure*}
\includegraphics[width=1.0\textwidth]{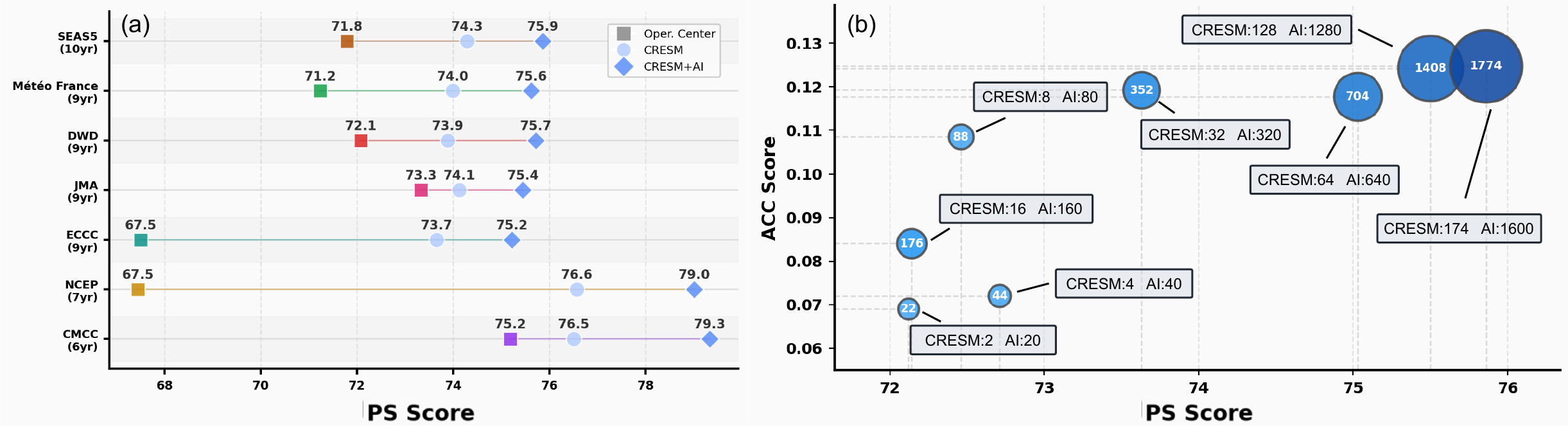}
% \vspace{-1.5em}
  \caption{Hindcast skill of CAPES and its ensemble-scaling behavior. (a) Comparison with state-of-the-art operational systems. (b) Forecast skill scaling with ensemble size under a fixed CRESM-to-AI ratio of 1:10, measured by PS and ACC.}
  \label{fig:center_comparison_scaling}
  % \vspace{-1.5em}
\end{figure*}

As shown in Fig.~\ref{fig:center_comparison_scaling}(a), the 174-member CRESM ensemble already surpasses the competing operational hindcasts, indicating strong intrinsic skill from the physically based forecasting component. After expanding the ensemble to 1,774 members with AI-generated forecasts, CAPES delivers a further improvement, demonstrating that scalable AI ensemble expansion can effectively enhance the prediction skill of the numerical core. These results confirm both the quality of the CRESM baseline and the value of large hybrid ensembles in seasonal precipitation forecasting. This demonstrates the strong baseline skill of CRESM, the effectiveness of AI-based ensemble expansion, and the benefit of large ensemble sizes.

Fig.~\ref{fig:center_comparison_scaling}(b) further shows a clear scaling relationship between ensemble size and forecast skill. We evaluate a series of hybrid ensembles with different numbers of CRESM and AI members, using a fixed 1:10 ratio. Both the 10-year mean PS and ACC increase systematically with ensemble size. In particular, the PS score rises from 72.1 for 22 members to 75.9 for the full 1,774-member ensemble, providing empirical evidence that forecast skill in CAPES follows a favorable ensemble-scaling law.

% \subsection{Operational Forecast for 2026}
% To demonstrate its practical applicability, we further use CAPES to forecast summer precipitation in 2026. The forecast results are shown in Fig.~\ref{fig:forecast_2026}, and was submitted to the flood-season forecasting consultation organized by the CMA, where it was considered together with other operational forecasts to support the official outlook for the 2026 flood season.

% \begin{figure}
% \includegraphics[width=\linewidth]{figures/CAPES_1774_2026.png}
% \vspace{-2em}
%   \caption{CAPES forecast of summer mean precipitation in 2026 with 1,774 ensemble members.}
%   \vspace{-2em}
%   \label{fig:forecast_2026}
% \end{figure}

\section{Implications}
% implications for future systems and applications (1 p max)
For flood-season forecasting, CAPES suggests that this long-standing problem can be pushed forward in a qualitatively different way. Over the 2016-2025 hindcast period, the full 1,774-member hybrid system improves the mean prediction score from ECMWF's 71.8 to 75.9 while also yielding more stable regional rainfall patterns and better coverage of extreme outcomes than existing systems.

\begin{figure}
\includegraphics[width=1\linewidth]{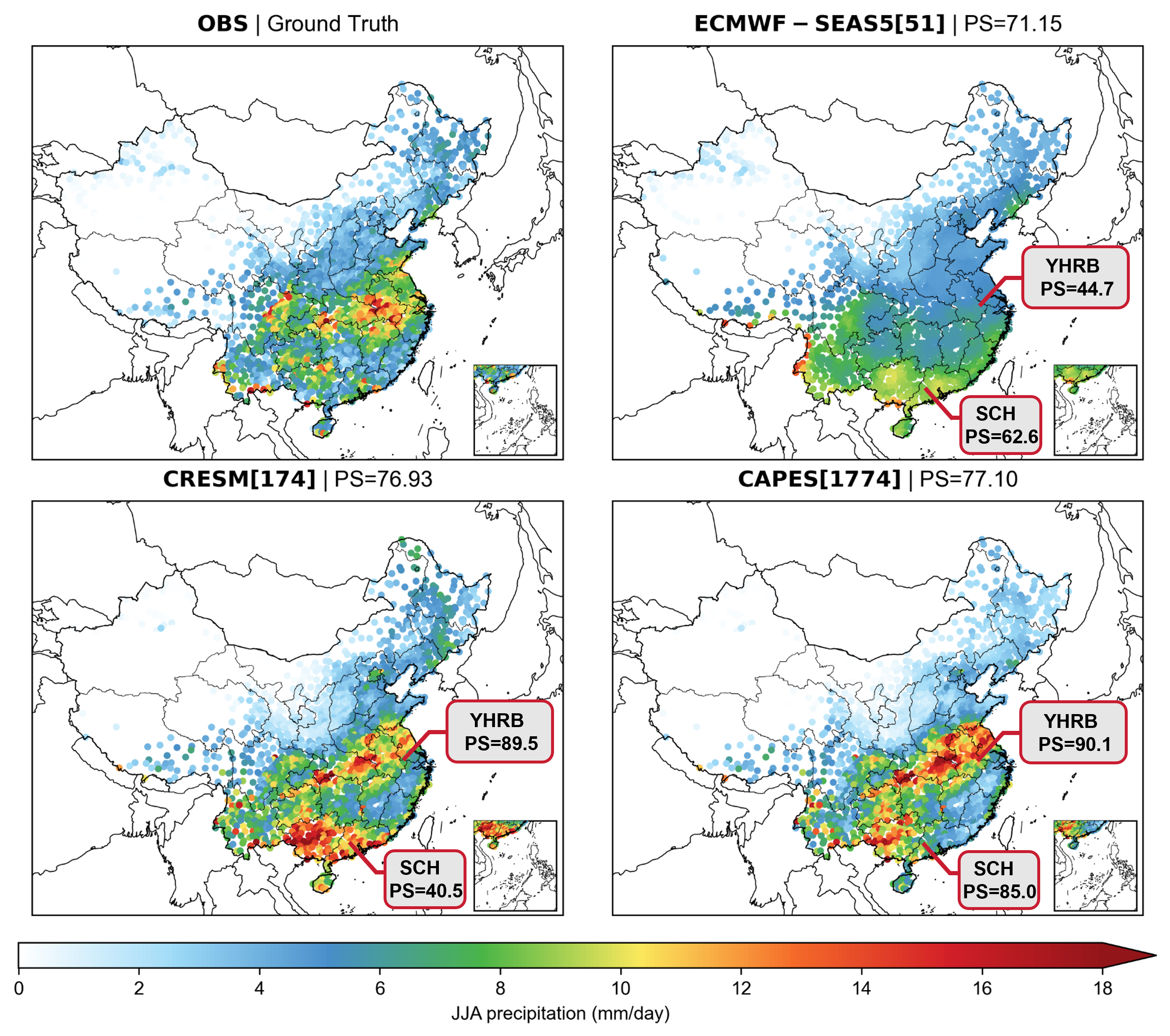}
\vspace{-2em}
  \caption{Comparison of summer precipitation hindcasts in 2020 from ECMWF, CRESM, and CAPES against observations.}
  \vspace{-2em}
  \label{fig:2020_comparison}
\end{figure}

As shown in Fig.~\ref{fig:2020_comparison}, the 2020 summer rainfall was concentrated around the Yangtze River basin and Southern China. ECMWF SEAS5 fails to reproduce the observed rainbelt structure, with clear underestimation over the Yangtze–Huaihe River basin (YHRB). CRESM captures the primary rainbelt and intensity over YHRB, with a PS score of 89.5, but exhibits evident spatial displacement over Southern China, leading to a reduction in PS score. In contrast, CAPES maintains the high skill over YHRB, reaching a PS score of 90.1, while largely correcting the spatial organization over Southern China, improving the score from 40.5 to 85.0, producing a more coherent rainfall pattern that better aligns with observations and more reliably represents extreme precipitation.

\begin{figure}
\includegraphics[width=1\linewidth]{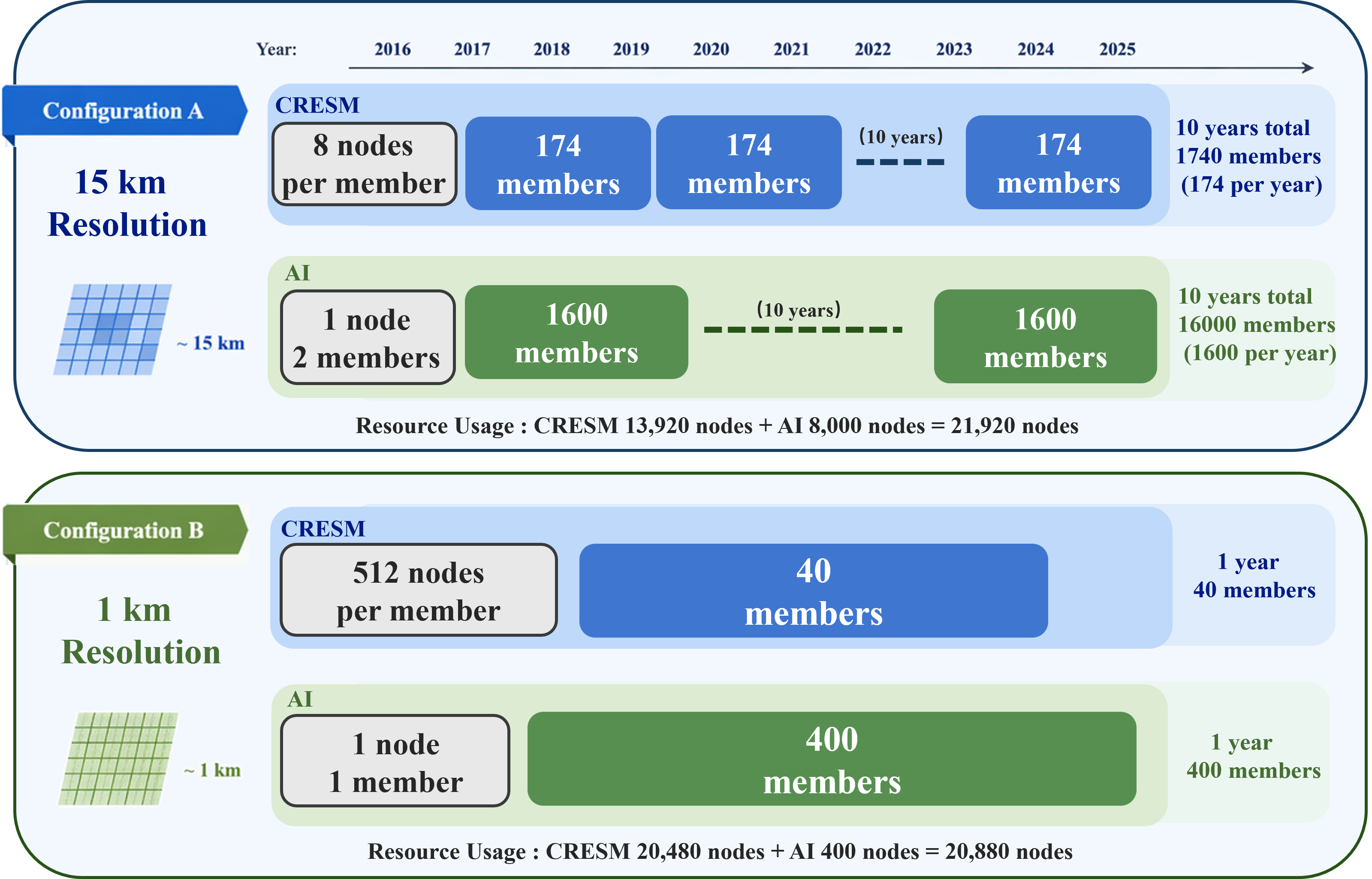}
\vspace{-2em}
  \caption{Concurrent full-machine execution modes of CAPES on LineShine for 15-km decadal hindcasts and 1-km operational forecasting.}
  \vspace{-2em}
  \label{fig:concurrency}
\end{figure}

The significance of this gain is not only that one forecast center is outperformed by another. More fundamentally, it shows that a central limitation of seasonal precipitation prediction, namely the persistent trade-off among physical fidelity, ensemble size, and operational cost, can be relaxed when a physically grounded numerical backbone, large AI ensemble expansion, and forecast fusion are designed as one workflow rather than as separate components.

Fig.~\ref{fig:concurrency} further shows the experimental and operational scale enabled by full-machine execution on LineShine. At 15-km resolution, one hybrid ensemble consists of 174 numerical members and 1,600 AI members. Each numerical member uses eight nodes and every two AI members share one node, so a complete ensemble occupies 2,192 nodes. Under full-machine operation, LineShine can execute ten such cases concurrently and finish the 2016-2025 hindcast campaign in 14.6 hours, making it possible to validate and iterate methods over a decadal sample within one day. At 1-km resolution, the full-machine limit supports one 440-member hybrid ensemble, including 40 numerical members and 400 AI members. Each 1-km CRESM member uses 512 nodes and each AI member uses one node, enabling completion of a six-month operational flood-season forecast in 158.8 hours, within about one week.

This shift also carries broader implications for Earth system science. By combining HPC and AI in one forecasting pipeline, we are not only improving seasonal prediction skill, but also changing how the problem can be studied. Numerical models continue to provide physically interpretable process constraints, explicit coupling pathways, and a foundation for mechanism-oriented reasoning, while AI makes it possible to absorb richer multi-source observations, represent uncertainty through much larger ensembles, and connect information across scales in ways that are difficult to realize in conventional workflows alone. The resulting hybrid system opens new routes to analyze uncertainty sources, probe potential causal chains, and gradually convert parts of the AI ``black box'' into more explainable scientific structure. At the same time, it alleviates one of the long-standing weaknesses of conventional numerical systems: their limited ability to efficiently exploit all available heterogeneous observations and rapidly evolve with new data streams and new scientific knowledge.

Seen from this perspective, the significance of CAPES extends beyond a single forecasting application. It offers an initial example of forecast-capability scaling: when physical modeling, AI, data, and ensemble design are jointly organized, gains in computing resources and data resources can be translated into measurable gains in operational prediction. More importantly, it points to a broader possibility for Earth system science: a co-evolution in which compute, data, theory, and models reinforce one another, so that the scaling of forecasting capability may be accompanied by a scaling of scientific discovery and disciplinary knowledge itself. In this view, larger compute is not valuable only because it runs larger simulations; it also provides the substrate for faster iteration between theory, data assimilation, model development, and AI. If such iteration becomes systematic, structured scientific knowledge extracted from hybrid forecasting systems may in turn enable theory-guided paradigm shifts, in a similar way to the case that deeper understanding of aerodynamic lift ultimately enabled the transition from observing bird flight to building jet aircraft.

Finally, LineShine illustrates the scientific value of a platform that natively supports both large-scale simulation and large-scale AI. Such an HPC-AI co-driven environment is not only important for climate and flood-season forecasting, but may also provide a general template for other domains that must combine structured numerical modeling with unstructured or heterogeneous data, including materials, biomedicine, and robotics. In this sense, the significance of the present work lies not only in the specific gains in skill and time-to-solution achieved on LineShine, but in showing a credible path by which future scientific software, workflows, and platforms can convert raw computing power into sustained advances in prediction, scientific understanding, and eventually theory-driven breakthroughs across disciplines.

\enlargethispage{1.08cm}  % 增大第一页下边距
\begin{acks}

%The authors thank \textbf{[Advisor Name]} and \textbf{[Collaborator Name 1]} for their guidance and insightful discussions.

This work is supported by \textbf{National Natural Science Foundation of China under Grant T2125006}. We acknowledge the computational resources provided by \textbf{National Supercomputing Center in Shenzhen}, which supported the large-scale experiments presented in this study. The integrated physical–AI framework involves substantial computational and system considerations, and benefited from co-design across models and implementation.

%We also thank \textbf{[Support Staff Name]} for their support in system operation and performance optimization.

\end{acks}

%%
%% The next two lines define the bibliography style to be used, and
%% the bibliography file.
\bibliographystyle{ACM-Reference-Format}
\bibliography{sample-base}
\end{document}